**The Causal Effect of Repealing Certificate-of-Need Laws for Ambulatory Surgical Centers: Does Access to Medical Services Increase?**

Thomas Stratmann
Department of Economics
George Mason University

Markus Bjoerkheim
Mercatus Center
George Mason University

Christopher Koopman
The Center for Growth and Opportunity
Utah State University

April 2024

In many states, certificate-of-need (CON) laws prevent ambulatory surgical centers (ASCs) from entering the market or expanding their services. This paper estimates the causal effects of state ASC-CON law repeal on the accessibility of medical services statewide, as well as for rural areas. Our findings show that CON law repeals increase ASCs per capita by 44-47% statewide and 92-112% in rural areas. Repealing ASC-CON laws causes a continuous increase in ASCs per capita, an effect which levels off ten years after repeal. Contrary to the "cream-skimming" hypothesis, we find no evidence that CON repeal is associated with hospital closures in rural areas. Rather, some regression models show that repeal is associated with fewer medical service reductions.

*JEL* Codes: I11, I18, L51, R51, K23

Keywords: certificate-of-need, certificate of need, certificate of need laws, healthcare, health care, regulation, entry barrier, barriers to entry, ambulatory surgical centers, community hospitals, rural health care, rural healthcare

* We thank two anonymous reviewers and the editor for their helpful comments and suggestions, and Ali Melad for excellent research assistance.

# 1. Introduction

Certificate-of-Need (CON) laws restrict entry and/or expansion of healthcare facilities in 35 states (National Conference of State Legislatures, 2024). These laws require hospitals, nursing homes, ambulatory surgical centers, and other healthcare providers to obtain regulatory approval before opening a new practice, expanding, or making certain capital investments. Thus, CON laws effectively create barriers to entry that limit competition among medical providers.

A substantial body of research has investigated whether CON laws successfully provide greater access to medical care, lower healthcare costs, and improve the quality of services.[1] In a literature review, Mitchell (2024, p. 1) finds overwhelming evidence that individuals residing in states with CON laws experience reduced access to medical care. Further, the medical care available in CON states is of higher cost and lower quality than states without CON laws. However, much of the current CON literature is limited by the use of empirical designs that can only speak to correlations between CON laws and the stated goals, without addressing causality. Exceptions include work by Cutler et al. (2010), Perry (2018), and Yu (2024). However, these studies analyze how CON repeals impact access to specific services in single states (Pennsylvania, North Carolina, and Missouri, respectively).[2]

This paper overcomes limitations in the previous CON literature by exploiting the staggered repeals of CON laws for Ambulatory Surgical Centers in six states on access to medical care, using an empirical design and estimation methods that allow for causal inference.

---

[1] The literature has studied how CON laws relate to medical costs, expenditures, quality, and provision for indigenous and underserved populations. Examples of this work are Sloan (1981); Joskow (1981); Robinson et al., (2001); Rivers et al., (2010); Stratmann and Russ (2014); Bailey (2016); Bailey (2018); Oshfeldt and Li (2018); Wu et al., (2019); Stratmann and Baker (2020); Baker and Stratmann (2021); Stratmann (2022). For a comprehensive review of the literature, see Mitchell (2024).

[2] The study closest to our approach is a contemporaneously developed paper by Melo, Sigaud, Neilson, and Bjoerkheim (2024, this issue) that examines the repeal of CON laws for hospital beds in five states. Another example is Chiu (2021), who estimates that the causal effect of CON laws is an increase in heart attack mortality, using a border discontinuity design.

We focus on ambulatory surgery centers due to their growing importance: In 1980, 16% of surgeries were performed on an outpatient basis. Hospitals provided most of these surgeries, as only a few hundred ASCs operated nationwide. Today, 80% of surgeries occur in outpatient settings, and the number of surgical centers has grown to almost 6,000, fueled by advancements in anesthesia and non-and minimally invasive surgical procedures (Hall, 2017; Munnich and Parente, 2018).

Ensuring access to healthcare in rural communities has remained a central justification of state CON legislation; however, this aspect has only received modest attention in the CON literature. Exceptions include Yu (2024), showing that the repeal of Missouri's ASC-CON increased ASC presence; Melo et al., (2024) showing that the repeal of CON for hospital beds caused hospital entry in both urban and rural areas, Carlson et al. (2010), showing that CON laws correlate with decreased rural access to hospice care; and Herb (2021), documenting that CON laws are associated with prolonged travel time to radiation oncology facilities in rural areas of the Northeast and Midwest, but not in the South. Our study contributes to the limited work on how CON laws affect access to rural medical services.

Hospitals and ASCs often compete in providing outpatient services (Carey et al., 2011; Plotzke and Courtemance, 2011). However, ASCs have more control over patient selection, as federal regulations permit surgical centers to selectively offer services based on subjective criteria, including whether a patient can be safely discharged following the procedure.[3] One often-cited rationale for restricting ASCs' entry is "cream-skimming," a catch-all phrase for providers accepting and providing services to only the most profitable patients (Rajan et al., 2021). CON advocates claim that limiting ASC entry reduces cream-skimming, assuring hospitals' viability and access to hospital medical care. We test the implications of the "cream-skimming" hypothesis, namely, whether repealing ASC-CON laws is associated with hospital closures or reductions in medical services.

---

[3] For example, if a patient does not have a responsible adult to transport the patient home and provide post-operative care, the anesthesiologist or surgeon may choose not to offer the surgery.

We show that repealing ASC-CONs causes a 44-47% increase in per capita ASC presence statewide and an increase of 92-112% in rural areas. The findings are robust across several modern difference-in-difference estimators and specifications. Further, our findings do not support the arguments of CON advocates who claim that CON laws reduce cream-skimming by ASCs, and thereby ensures access to hospital services by preventing hospital closures. Rather, the estimates show that states that repeal CON laws do not have more hospital closures, but rather, have fewer hospital service reductions than states with CON laws. These findings suggest that repealing ASC entry restrictions may instead facilitate the survival of rural hospitals.

## 2. A Brief History of Certificate-of-Need, Rural Access, and Cream-Skimming

In 1964, New York became the first state to pass CON legislation. The legislation aimed to strengthen regional health planning programs by creating a process for prior approval of certain capital investments (Simpson, 1985). Between 1964 and 1974, 26 other states adopted CON legislation. However, with the passage of the National Health Planning and Resources Development Act of 1974 (NHPRDA), the federal government made the availability of some federal funds contingent on the enactment of state CON legislation. By the end of 1982, every state except Louisiana had passed a CON law regulating hospitals, nursing homes, dialysis facilities, and ambulatory surgical centers (Simpson, 1985).

However, evidence accumulated that CON laws were failing to achieve their goals (Cimasi, 2005). Several states, including Texas, Arizona, and Utah, repealed their CON laws.

Federal legislators became increasingly concerned that CON laws had "failed to reduce the nation's aggregate healthcare costs, [and were] beginning to produce a detrimental effect in local communities." (McGinley, 1995; Simpson, 1985). In 1986, Congress repealed the NHPRDA, ending the federal government's subsidization of state CON laws.[4]

After the NHPRDA's repeal, several states repealed their CON laws. By the end of the 1980s, twelve states had eliminated their CON laws (Arizona, California, Colorado, Idaho, Kansas, Minnesota, New Mexico, South Dakota, Texas, Utah, Wisconsin, and Wyoming). Between 1990 and 2000, three states followed (Indiana, North Dakota, and Pennsylvania).[5] Since 2000, Wisconsin, New Hampshire, Florida, South Carolina, and Montana have eliminated all or most of their CON laws.[6]

CON laws have been justified on the grounds of achieving numerous public policy goals. Policymakers have seen CON laws as a way for governments to control healthcare costs, regulate the level of capital investments, increase charity care, protect the quality of medical services, and protect access to services across geographic locations. Indeed, Congress cited rural access to medical care as one of the primary goals of the 1974 National Health Planning and

---

[4] For an in-depth discussion of the NHPRDA, see Madden (1999).
[5] Indiana reinstated its CON law in 2018.
[6] Wisconsin has repealed its CON law twice. South Carolina repealed its CON law in 2023, but the laws are in effect until 2027. After repealing most CON laws, Florida and Montana maintained their CON laws for long-term care facilities.

Resources Development Act. After federal repeal, many states adopted rural access to medical services as a primary rationale for continuing to implement CON laws.[7,8,9]

For example, Pennsylvania's CON laws required the "identification of the clinically related health services necessary to serve the health needs of the population of this Commonwealth, including those medically underserved areas in rural and inner-city locations." The North Carolina CON statute states that "access to healthcare services and healthcare facilities is critical to the welfare of rural North Carolinians, and to the continued viability of rural communities, and that the needs of rural North Carolinians should be considered in the certificate of need review process." A stated goal of Virginia's CON law is to support the "geographical distribution of medical facilities and to promote the availability and accessibility of proven technologies." And one of the justifications for West Virginia's CON laws is that they provide "some protection for small rural hospitals . . . by ensuring the availability and accessibility of services and to some extent the financial viability of the facility."[10]

---

[7] The NHPRDA included National Health Priorities, which begin with the goal of "the provision of primary care services for medically underserved populations, especially those which are located in rural or economically depressed areas."

[8] See, e.g., Arkansas (A.C.A. § 20-8-103(b)-(c)); Florida (Fla. Stat. Ann. § 408.034(3)); Georgia (Ga. Code Ann., § 31-6-1); Kentucky (KRS § 216B.010); North Carolina (N.C. Gen. Stat. Ann. § 131E-175(3a)); Tennessee (Tenn. Code Ann. § 68-11-1625(c)(7)); Virginia (12 Va. Admin. Code 5-230-30(2)), 12 Va. Admin. Code 5-230-30(2) (2015), 35 Pa. Stat. § 448.401c

[9] To understand the theoretical underpinnings for using CON laws to protect access, see Colon Health Centers of America v. Hazel et al., No. 14-2283, slip op. at 23 (4th Cir. 2016), which notes,
> A related purpose of the CON law is geographical in nature. For reasons not difficult to discern, medical services tend to gravitate toward more affluent communities. The CON law aims to mitigate that trend by incentivizing healthcare providers willing to set up shop in underserved or disadvantaged areas such as Virginia's Eastern Shore and far Southwest. "In determining whether" to issue a certificate, for example, Virginia considers "the effects that the proposed service or facility will have on access to needed services in areas having distinct and unique geographic, socioeconomic, cultural, transportation, or other barriers to access to care." Va. Code Ann. § 32.1-102.3(B)(1).

A CON law may also aid underserved consumers more indirectly. By reducing competition in highly profitable operations, the program may provide existing hospitals with the revenue they need not only to provide indigents with care but also to support money-losing but important operations like trauma centers and neonatal intensive care units.

[10] West Virginia Health Care Authority, *Annual Report to the Legislature 1998*, http://www.hca.wv.gov/data/Reports/Documents/annualRpt98.pdf.

To achieve greater access to medical care in rural communities, many states use CON laws to limit entry and expansion of medical providers, including firms deemed "hospital substitutes," such as ASCs (Cimasi, 2005). ASCs are subject to federal regulations permitting them to treat patients only if the service is not expected to exceed 24 hours and does not require subsequent hospitalization.[11] Regulators consider limiting ASC entry beneficial, as these providers allegedly engage in "cream-skimming," meaning that surgery centers treat only the most profitable patients, thereby reducing hospitals' profit centers. Some CON advocates claim that when more profitable patients seek care elsewhere, hospitals' ability to cross-subsidize charity care and provide other essential services is reduced, potentially resulting in hospital closures.

Scholarly work has researched cream-skimming by ASCs (Plotzke and Courtemance 2011; Munnich and Parente 2018), hospitals (Friesner and Rosenman 2009; Yang et al., 2020), and outside of healthcare settings (Tabarrok 2013). However, the cream-skimming hypothesis has not been tested in the context of the ASC-CON repeal.

## 3. Hypotheses

While CON laws can influence healthcare markets along several margins, we first test whether ASC-specific CON laws act as barriers to entry, given that these laws aim to reduce the number of ASCs in a state.[12] If CON laws are barriers to entry, we predict their repeal will lead to increased ASCs per capita operating in the state. Given the explicit rationale for CON laws to provide access to medical care in rural areas, we add the hypothesis that repealing ASC-CONs results in more ASCs in rural areas.

---

[11] 42 CFR Part 416, available at https://www.ecfr.gov/current/title-42/chapter-IV/subchapter-B/part-416.
[12] CON laws are also barriers to expanding existing facilities, not just to the entry of new facilities. In this paper, we do not analyze this aspect of ASC CON laws.

*Hypothesis 1:* *Repealing ASC-CON laws increases ASCs per capita statewide.*

*Hypothesis 2:* *Repealing ASC-CON laws increases ASCs per capita in rural areas.*

CON advocates claim that limits to ASC entry reduce cream-skimming, which protects the viability of incumbent hospitals in rural areas and prevents them from closing. Similar reasoning predicts that this prevents rural hospitals from reducing the services they offer, what's known as a conversion. Examples include hospitals that close their inpatient units but continue to operate at a reduced capacity, converting to standalone emergency departments, outpatient care centers, or specialized medical facilities.

*Hypothesis 3:* *Repealing an ASC-CON law increases hospital closures or service reductions by hospitals in rural areas.*

## 4. Data and Empirical Strategy

### 4.1. Ambulatory Surgical Centers

Our data source for ASCs is the POS files from the Centers for Medicare and Medicaid Services accessed through the National Bureau of Economic Research (NBER).[13] We construct two state-level annual measures: the number of operating ASCs per 100,000 state population and the number of operating ASCs per 100,000 rural population from 1991-2019. We merge the POS data with the 2013 edition of the Urban-Rural classifications from the National Center for Health

---

[13] The data in this paper builds on Stratmann and Koopman (2016).

Statistics (NCHS) to determine whether an ASC is in an urban or rural community. We classify providers as rural if they are located in a Micropolitan or Non-Core area.[14]

Data on Certificate-of-Need laws come from the American Health Planning Association (AHPA) and HeinOnline's Digital Session Laws Library. Six states repealed ASC-CON laws from 1991-2019: Pennsylvania (1996), Ohio (1997), Nebraska (1999), New Jersey (2000), Missouri (2002), and New Hampshire (2016).[15]

Our empirical design uses a difference-in-difference approach: the treatment group comprises the six repeal states, and the states with CON laws throughout our sample period comprise the control group.[16] This approach estimates the causal effect of ASC-CON repeals, assuming parallel trends between the treatment and control groups. The difference-in-difference regression model for the Two-Way Fixed Effects model (TWFE) model is

$$ASCs\ per\ 100{,}000_{it} = \gamma\ ASC\text{–}CON\ Repeal_{it} + \mu_t + \alpha_i + \varepsilon_{it} \quad (1)$$

and the regression model for the corresponding event study is

$$ASCs\ per\ 100{,}000_{it} = \sum_t \lambda_t\ ASC\text{–}CON\ Repeal_{it} + \mu_t + \alpha_i + \eta_{it} \quad (2)$$

In Equation (1), we estimate the average impact of repealing ASC-CON on the number of ASCs per 100,000 population in state *i* and year *t*. Here, the CON variable is an indicator variable equal to one starting in the first year a state's CON repeal is in effect and zero otherwise. We include year effects $\mu_t$ and state effects $\alpha_i$. We estimate both equations for ASCs statewide and for ASCs in a state's rural areas. We cluster the standard errors by state throughout.

---

[14] The Urban-Rural classifications change several times during our sample period. We use the 2013 edition to ensure our unit of analysis remained constant over our study period (Ingram and Franco, 2014). This circumvents the possibility that changes to which counties are classified as Urban or Rural in the different editions could introduce compositional changes to our panel. We were able to link all providers to Urban-Rural codes after harmonizing the FIPS codes of a small number of counties using https://seer.cancer.gov/popdata/modifications.html#appendix1 and by using the ZIP codes for a handful of providers with missing FIPS codes in the POS data. No rural counties exist in one of the repeal states (New Jersey) and three of our control states (Delaware, Rhode Island, and the District of Columbia). Rural analyses are, therefore, based on a slightly smaller sample.
[15] Pennsylvania and New Hampshire eliminated their entire CON laws.
[16] We do not include California, Colorado, Idaho, Kansas, North Dakota, Texas, Utah, Wyoming, and South Dakota from the control group, as they did not maintain a CON law throughout our sample period. We include the District of Columbia in the control group for statewide analyses, but not when we analyze rural areas, as described above. We did not include Indiana in the control group as it repealed and implemented CON several times during our sample period. As a robustness test, we present results in the appendix, using all non-repeal states as the comparison group.

In Equation (2), we implement an event-study approach. That model interacts with the repeal indicator with year indicators before and after CON repeals, centered around the final year before each repeal's implementation. We plot the estimated coefficients for ten years before and 20 years after the repeal. This allows us to assess the dynamic effects of repeal and evaluate the plausibility of the assumption of parallel trends.

Recent econometric developments have identified several biases when attempting to estimate treatment effects using Equations (1) and (2) when the timing of treatment differs across treated units and when treatment effects are heterogeneous across time or units (Borusyak, Jaravel, and Spiess, 2021; Callaway and Sant'Anna, 2021; Chaisemartin and D'Haultfœuille, 2020; Gardner, 2021; Wooldridge, 2021). To obtain unbiased estimates of Models (1) and (2), we employ the "Extended-TWFE" approach proposed by Wooldridge (2021), the imputation-based approach proposed by Borusyak, Jaravel, and Spiess (2021), estimators developed by Callaway and Sant'Anna (2021), the two-stage approach of Gardner (2021), and the estimator by de Chaisemartin and D'Haultfœuille (2020).

We use a pooled reference period for the Wooldridge estimator, which generates more precise estimates than a fixed reference period.[17] We implement Borusyak's imputation method with a 10^-6 default tolerance and 100 iterations. The Callaway-Sant'Anna method can be estimated with the control group either being states that never repealed or states that have "not yet" repealed. We present both approaches. Gardner's two-stage procedure first estimates state and year-fixed effects using observations from states that did not or have not yet repealed CON to predict the counterfactual outcomes for all units and periods. The residualized outcome is then regressed on repeal in the second stage to estimate the effect of repeal. We estimate de Chaisemartin and D'Haultfœuille's proposed method with the first-difference option, comparing first-time treated units to placebo's not-yet treated. We draw 100 bootstrap replications to compute the standard errors clustered on the state level.

---

[17] For further discussions see Wooldridge (2021) and Hegland (2023).

## 4.2. Rural Hospital Closures and Hospital Medical Service Reductions

To test the hypotheses regarding rural hospital closures and service reductions, we use the Rural Hospital Closures data from the University of North Carolina (UNC Cecil G. Sheps Center for Health Services Research, 2024). This data set includes measures of reductions in rural healthcare access, overcoming limitations in the POS data (Kaufman et al., 2016). The UNC data classify a hospital as closed if it ceased providing general, short-term, and acute inpatient care. The data classify a hospital as converted if it closed its inpatient unit but continued to offer some medical services.[18] Converted hospitals are hospitals with service reductions.

The beginning year of the UNC data is 2005. All repeal states other than New Hampshire (2016) removed ASC-CONs from their statutes prior to 2005, making the use of a difference-in-difference empirical design unfeasible. Instead, we compare the repeal states to states with CON laws on four measures of reductions in healthcare access, from 2005 to 2019. The measures are rural hospital closures, service reductions, beds closed, and beds involved in service reductions per 100,000 rural population. There are no rural counties in three CON states (Delaware, Rhode Island, and the District of Columbia) and one repeal state (New Jersey). The rural analyses are, therefore, based on this slightly smaller sample. The regression model is:

$$Y_i = \alpha + \gamma\, ASC\text{-}CON\ Repeal_i + \boldsymbol{\beta X_i} + \varepsilon_i \qquad (3)$$

where $Y_i$ is one of the outcome measures. The matrix **X** includes the control variables: rural population as a percentage of the state in 2005, the average unemployment rate from 2005 to 2019, and the percent change in the rural, Black, Hispanic, and elderly populations between 2005 and 2019. To control for changes to residents' health status, we include the percent change in mortality rates due to lung cancer or diabetes for residents 18 and older between 2005 and 2016. The CDC ended the construction of these series in 2016; thus, 2016 is the last year of this variable. The Census Bureau is the data source for population size, rural population size, and percentages of black, Hispanic, and elderly populations, defined as those 65 and older. State-level data on unemployment rates come from the Bureau of Labor Statistics.

---

[18] This data is only available for rural areas. The data does not consider a hospital to have closed if it merged with another hospital or was sold to another hospital and continued operating. Hospitals that converted to Critical Access Hospitals, Rural Emergency Hospitals, or which closed and reopened within the same calendar year at the same location are also not considered closed, following the methodology developed by the Office of Inspector General (Rehnquist, 2003).

## 5. Results

Table 1, Panel A, shows summary statistics for these variables from 1991-2019. Table 1, Panel B shows summary statistics for the states that repealed ASC-CON, as well as for states with CON laws. Table 1, Panel C shows descriptive statistics for the measures we use to analyze hospital closures and medical service reductions. These are presented separately for the states that repealed ASC-CONs and states that had CON laws over the sample period. Panel C foreshadows some of our estimation results by documenting that repeal states experience fewer hospital closures and service reductions from 2005-2019 than did CON states. Panel C also shows that closed hospitals are similar in repeal states and CON states; however, hospitals that reduced services were smaller in repeal states.

**5.1. Ambulatory Surgical Centers**

Figure 2 plots the event-study coefficients from the difference-in-difference estimates from the regression models testing the effect of ASC-CON repeals on the number of ASCs statewide. In this Figure, the whiskers indicate the 95 percent confidence intervals corresponding to each point estimate. Leading up to date of the repeals, all point estimates of the unbiased estimation methods are closely centered around zero, supporting the parallel trends assumption. The point estimates are positive after repeal, statistically significant after three to four years and increase steadily during the first ten years. These findings indicate that ASC-CON repeals cause ASC entry. About ten years after the repeal, the increase in ASCs levels off at approximately 0.75 additional ASCs per 100,000 state population.

In Figure 2, in contrast to the results documented by the unbiased estimators, the biased TWFE-estimator estimates the presence of pre-trends and somewhat smaller point estimates following repeal. Given that this biased estimator detects pre-trend trends, if one were to account for these by estimating linear trends before treatment and subtract the estimated trend from observations after treatment, the TWFE-estimator would estimate coefficients that would lead to the erroneous conclusion that CON repeal did not affect ASC entry (Rambachan and Roth, 2023).

Table 2 shows the corresponding estimates of the average treatment effect on the treated (ATT). The ATT estimates are of similar magnitudes, ranging from 0.522-0.553, and are statistically significant at the 1% and 0.1 % levels. These results imply that the repeal of ASC-CONs caused an increase in the number of ASCs per capita of 44-47% relative to baseline (1.179) shown in Table 1, Panel D. Combined with our event-study results, these estimates support our hypothesis that repealing ASC-CONs causes the entry of ASCs statewide.

**5.2. Rural Ambulatory Surgical Centers**

Figure 3 shows the event study plots documenting the effect of ASC-CON repeal on the entry of ASCs in rural areas. Pre-repeal and with two exceptions, the point estimates are centered closely around zero, supporting the parallel-trends assumption. The exceptions are the biased TWFE estimator and the Wooldridge extended TWFE estimator. The pre-trends estimates are noisy for the latter estimator, especially for the early pre-treatment years. After repeal, the point estimates for most estimators are positive and statistically significant after about four years. The estimated coefficients grow steadily during the first ten years after ASC-CON repeal and then level off at approximately 0.7-0.8 additional ASCs per 100,000 rural population.

Table 3 shows the estimated ATT effects corresponding to Figure 2. For the unbiased estimators, these estimates are positive, ranging from 0.499-0.610, and statistically significant. In comparison, the estimated ATT derived from the biased TWFE estimator is not statistically significant.

Similar to Table 2, the results for ASC entry in rural areas show that removing an ASC-CON increases the number of Ambulatory Surgical Centers in these areas. These estimates show that the number of ASCs in rural areas increases by 92-112% relative to the baseline (0.539) shown in Table 1, Panel D. Together with our event-study results, we find support for our hypothesis that repealing ASC-CONs causes ASCs to enter rural areas.

**5.3. Robustness Tests for the Analyses of ASC Entry: A Different Control Group**

As a robustness test, we re-estimate all models with both states that never had a CON and states that had a CON throughout the sample period in the control group. These results are presented in the Appendix, Figures A1 and A2, and Tables A1 and A2.

The estimates from these robustness tests are similar to those using CON states as the control group. The event studies support the parallel trends assumption, grow over time, and then level off about ten years after repeal. All estimated ATTs are statistically significant for both statewide and rural areas. As predicted, the magnitudes are slightly smaller when non-CON states are included in the control group, as entry is not restricted in these states.

**5.4. Hospital Closures and Service Reductions**

Table 4 presents the results from estimating Equation (3) with hospital closures per 100,000 rural population as the dependent variable. The estimated coefficient in Table 4, Column (1), is negative, suggesting that ASC growth is not associated with hospital closures. The magnitude of this estimate shows that ASC-CON repeal states exhibit, on average, 0.052 fewer rural hospital closures; however, this estimate is not statistically significant. In Table 4, all coefficients on hospital closures are negative, ranging from -0.052 in Column (1) to -0.007 in Column (3). While these estimates are not statistically significant, they do not support the hypothesis that repeal states lost access to rural hospital services, as measured by hospital closures.

Table 5 presents the results from estimating Equation (3) with closed hospital beds per 100,000 rural population as the dependent variable. In all specifications, the estimated coefficients are positive and range from 0.730-2.393; however, none of the point estimates are statistically significant. These results suggest that the rural per capita number of hospital beds closed is similar among repeal states and CON states. The results do not support the hypothesis that repeal states experience more closed hospital beds.

Table 6 presents the results from Equation (3), with hospital service reductions per 100,000 rural population as the dependent variable. The point estimate in Table 6, Column (1) shows that repeal states experienced 0.091 fewer service reductions by rural hospitals. This estimated coefficient is statistically significant at the ten percent level. The -0.085 estimate in Table 6, Column (2) is statistically significant at the ten percent level. The estimated coefficients on ASC-CON repeals in the subsequent columns are also negative and of similar magnitude, ranging from 0.070 to -0.107; however, they are not statistically significant. These results do not support the hypothesis that repeal states lost access to rural hospital services, as measured by hospital conversions. Rather, the evidence points towards ASC-CON repeal states having fewer reductions in hospital services.

Table 7 presents regression results from estimation Equation (3). In these regressions, the dependent variable is hospital beds lost due to hospital service reductions, measured as the number of beds lost per 100,000 rural population. The coefficient in Table 7, Column (1), shows the estimate that when no control variables are included, repeal states lost on average 7.3 fewer beds. This coefficient is statistically significant at the five percent level. In Columns (2)-(5), which add different control variables in each specification, the estimated coefficients range from -12.3 to -6.9. The coefficients in Column (3) and (5) are statistically significant at the ten percent level. Thus, evidence shows that fewer beds were lost in repeal states than in CON states when hospitals reduce services. These results do not support the hypothesis that repeal states lost access to hospital services, as measured by the size of the hospitals that reduced services. Instead, the evidence shows repeal states had smaller hospitals involved in service reductions.

## 5.6. Robustness Tests for the Analyses on Hospital Closures/Service Reductions

To evaluate the sensitivity of the results for hospital closures and service reductions we implement the same robustness test as in Section 5.3, by re-estimate all models with both states that never had a CON and states that had a CON throughout the sample period in the control group. These results are presented in the Appendix, Tables A4-A7. The results found in these robustness tests are qualitatively very similar to those using CON states as the control group. The estimated coefficients are similar throughout, with some minor differences found in the precision of the estimates, specifically in Tables A6 and A7.

The estimated coefficients in Columns (1) and (2) of Table A6 are of similar magnitude as before, but are now statistically significant at the five percent level, rather than the ten percent level (in Table 6). The coefficient in Column (3) is statistically significant at the ten percent level in Table A6. The finding that repeal states appear to have experienced fewer hospital service reductions (Table 6) is thus found to be robust to this alternative control group.

The estimated coefficients in Table A7 are of similar magnitude but Column (3) is statistically significant at the five percent level (rather than the ten percent level in Table 7), and the coefficient in Column (3) is statistically significant at the five percent level (rather than at the ten percent level). The finding that repeal states appear to have lost fewer hospital beds to service reductions (Table 7) is thus found to be robust to this alternative control group.

## 6. Conclusions

This study estimates the causal effect of repealing ASC-specific CON laws. It shows that ASCs per capita increased by 44-47% statewide due to repeal. In rural areas, ASC-CON repeal caused ASC's per capita to increase of 92-112%. Given that CON law repeal is followed by medical provider entry, these findings document that CON laws are effective statewide and rural entry barriers.

According to the cream-skimming hypothesis, unrestricted entry for hospital substitutes, such as ASCs, allows entrants to selectively provide services to the most profitable patients, thereby threatening hospitals' financial prospects. While our models cannot test this claim directly, they test the implications stemming from the cream-skimming hypothesis. Specifically, this hypothesis predicts increased hospital closures and service reductions in states that repeal their CON laws. We test this prediction for rural areas in CON repeal states, as closures of hospitals in rural areas are a major concern of policymakers, who favor CON laws based on the claim that they prevent hospital closures in rural areas. However, the point estimates on CON repeal do not support the prediction that ASC entry results hospital closures or hospital service reductions. Rather, we find suggestive evidence that repealing ASC-CONs improves access to hospital services.

One explanation for this suggestive evidence is that ASCs and hospitals serve complementary roles in healthcare markets. ASC entry allows hospitals to focus on surgeries and medical services that are not feasible in ASC settings. This differentiation could lead hospitals to specialize in more complex—and potentially more profitable—surgeries and make the hospital attractive for medical providers specializing in these services. Another explanation for our finding is that unrestricted ASC entry mitigates one reason hospitals close, e.g., the lack of qualified staff. Surgeons tempted to quit their rural hospital jobs are incentivized to continue working at the hospital when they can also form an ASC, 91% of which are surgeon-owned (Munnich and Parente, 2017). This explanation is also consistent with research that shows surgeons are the only physician-specialty that predicts rural hospital closures (Germack et al., 2019).

**Tables and Figures:**

**Table 1, Panel A. Summary Statistics from State Annual Data, 1991–2019**

|  | (1) Mean | (2) Std. dev. | (4) Min | (5) Max |
|---|---|---|---|---|
| ASCs per 100K State Population | 1.266 | 0.966 | 0.125 | 6.312 |
| Rural ASCs per 100K Rural Population | 0.692 | 0.632 | 0.000 | 3.978 |
| Black % State Population | 0.135 | 0.120 | 0.003 | 0.663 |
| Hispanic % State Population | 0.081 | 0.085 | 0.004 | 0.495 |
| Elderly (65+) % State Population | 0.137 | 0.022 | 0.041 | 0.212 |
| Unemployment Rate | 5.635 | 1.868 | 2.100 | 13.800 |
| N (N Rural) | 1,189 | | (1,073) | |

Note: There are fewer state-year observations in rural areas. Delaware, New Jersey, Rhode Island, and the District of Columbia do not have rural counties.

**Table 1, Panel B. Summary Statistics ASC-CON Repeal States and CON States**

|  | ASC-CON Repeal States | | CON States | |
|---|---|---|---|---|
|  | (1) Mean | (2) Std. dev. | (3) Mean | (4) Std. dev. |
| ASCs per 100K State Population | 1.454 | 0.725 | 1.234 | 0.977 |
| Rural ASCs per 100K Rural Population | 0.905 | 0.632 | 0.744 | 0.634 |
| Black % State Population | 0.095 | 0.049 | 0.142 | 0.127 |
| Hispanic % State Population | 0.058 | 0.051 | 0.085 | 0.089 |
| Elderly (65+) % State Population | 0.143 | 0.015 | 0.136 | 0.023 |
| Unemployment Rate | 5.115 | 1.778 | 5.725 | 1.870 |
| N (N Rural) | 174 (145) | | 1,015 (928) | |

Note: There are fewer state-year observations in rural areas. Delaware, New Jersey, Rhode Island, and the District of Columbia do not have rural counties.

**Table 1, Panel C. Summary Statistics Rural Hospital Closures and Service Reductions, 2005–2019**

|  | ASC-CON Repeal States | | CON States | |
|---|---|---|---|---|
|  | (1) Mean | (2) Std. dev. | (3) Mean | (4) Std. dev |
| Hospital Closures per 100K | 0.105 | 0.150 | 0.156 | 0.187 |
| Closed Hospital Beds per 100K | 5.376 | 8.083 | 4.646 | 6.535 |
| Hospital Service Reductions per 100K | 0.086 | 0.058 | 0.177 | 0.230 |
| Hospital Size in Service Reductions per 100K | 1.520 | 1.061 | 8.852 | 18.083 |
| Rural Population, Baseline | 0.271 | 0.113 | 0.253 | 0.176 |
| Unemployment Rate | 5.180 | 1.297 | 5.916 | 0.985 |
| Rural Population, % Change | -0.012 | 0.026 | 0.022 | 0.063 |
| Elderly (65+), % Change | 0.384 | 0.130 | 0.517 | 0.182 |
| Hispanic, % Change | 0.725 | 0.097 | 0.627 | 0.189 |
| Black, % Change | 0.325 | 0.288 | 0.340 | 0.275 |
| Adults Diagnosed with Diabetes + Lung Cancer Deaths per 100K (18+, Age-Adjusted), % Change | 0.942 | 0.057 | 0.913 | 0.073 |
| N | 5 | | 32 | |

Note: This table presents summary statistics for data used to analyze rural hospital closures, service reductions, and control variables. Therefore, New Jersey, Delaware, Rhode Island, and the District of Columbia do not have rural areas and are omitted. The variable Rural Population is the percent of the state population that was rural in 2005. The unemployment rate is averaged over the 2005-2019 period. The percent changes in the rural, Elderly (65+), Black, and Hispanic populations are calculated from 2005 to 2019. The percent change in adults (18+) diagnosed with diabetes and lung cancer is calculated from 2005-2016, as these series were discontinued in 2016.

**Table 1, Panel D. Ambulatory Surgical Centers at Baseline in ASC-CON Repeal States**

|  | (1) Mean | (2) Std. dev. |
|---|---|---|
| ASCs per 100K State Population | 1.179 | 0.529 |
| Rural ASCs per 100K Rural Population | 0.539 | 0.432 |
| N (N Rural) | 6 (5) | |

Note: This table presents the average number of Ambulatory Surgical Centers per 100,000 in the last year before ASC-CON repeal in repeal states. There are fewer observations in rural areas because New Jersey does not have rural counties.

**Figure 1. Certificate-of-Need Repeals for Ambulatory Surgical Centers (ASCs)**

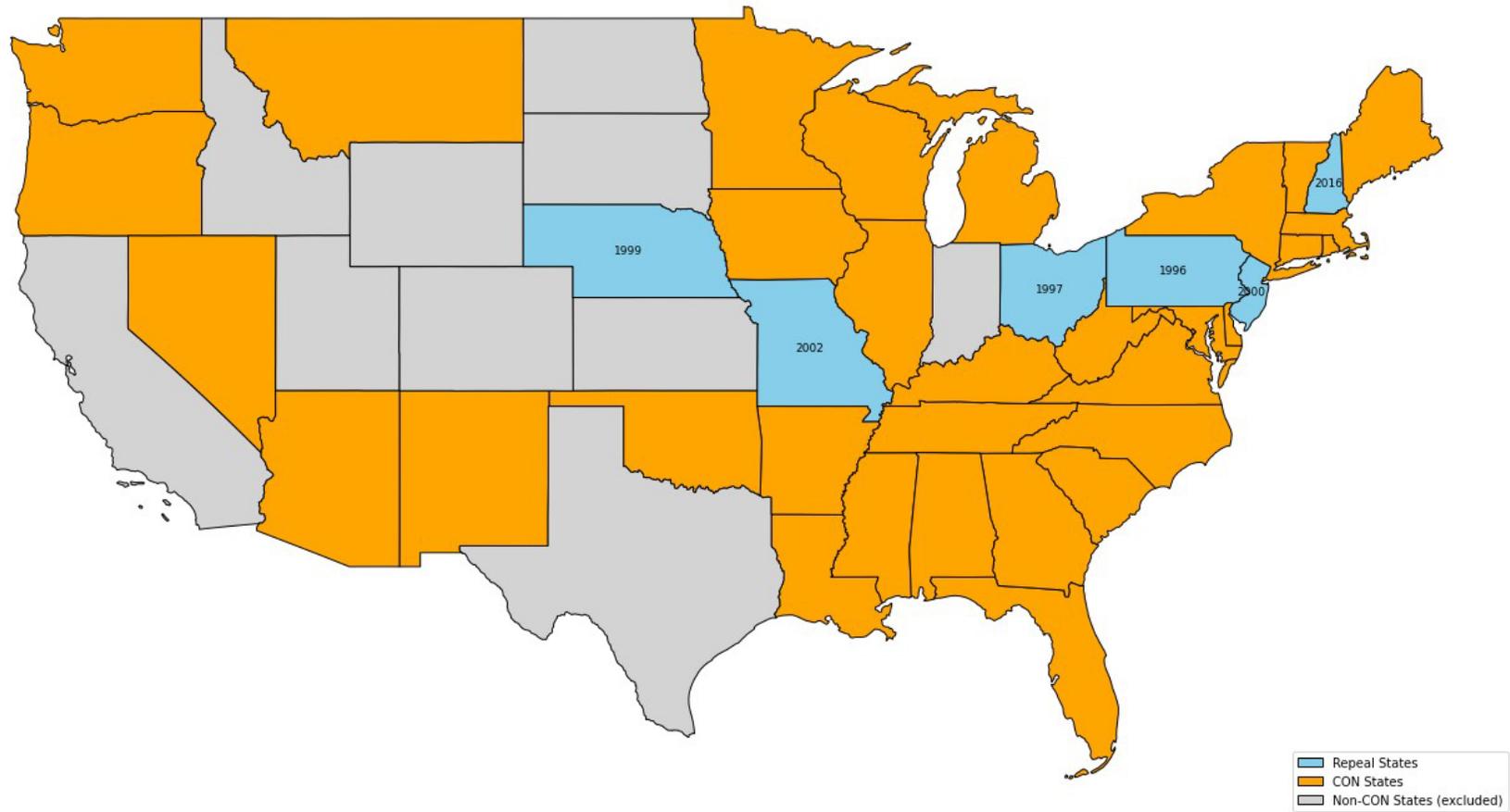

Source: https://www.mercatus.org/publication/district-columbia-and-certificate-need-programs-2020

**Figure 2. Statewide Event-Study Results: Impact of ASC-CON Repeal on Ambulatory Surgical Centers per 100,000 State Population**

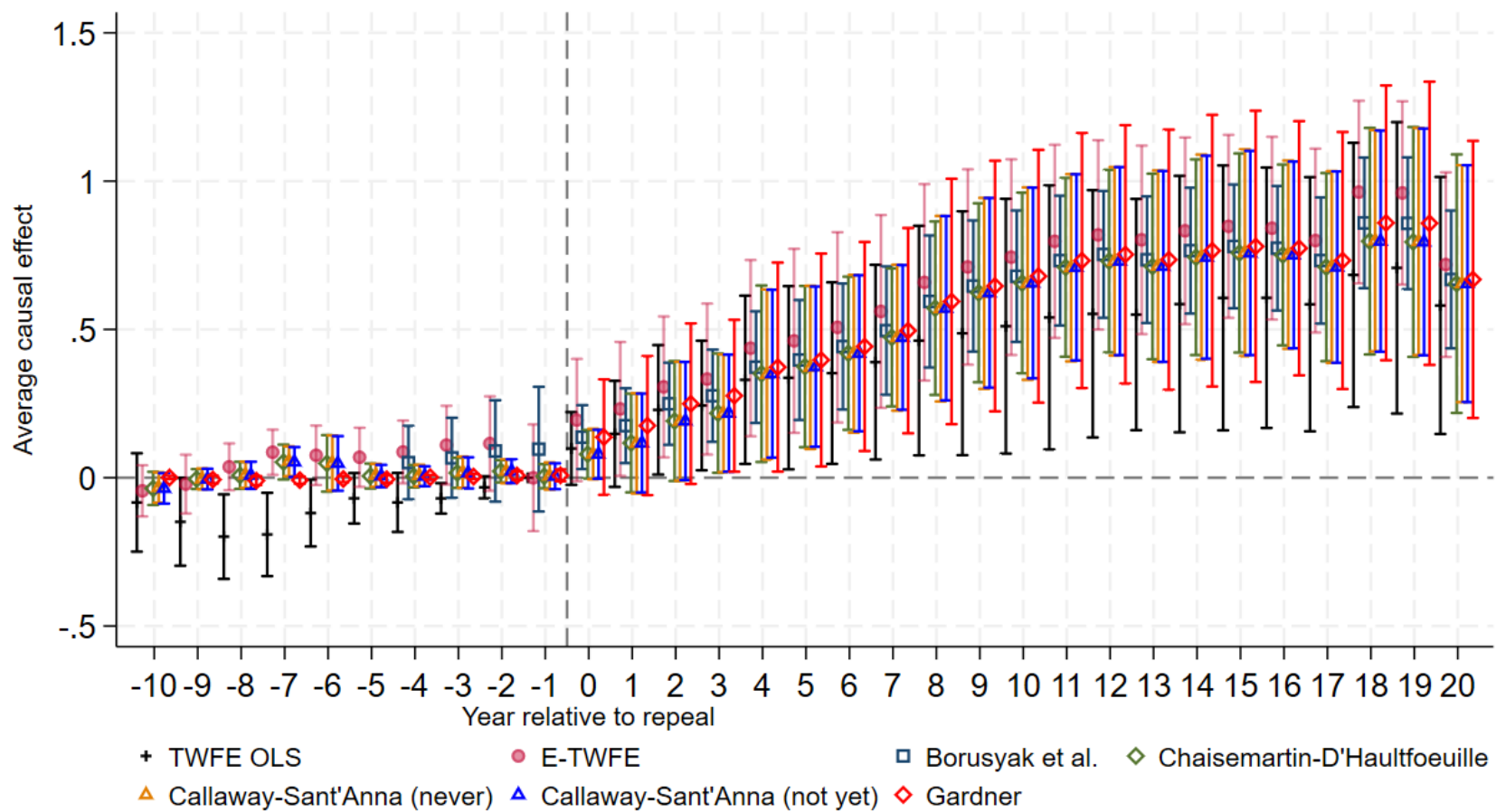

**Figure 3. Rural Event-Study Results: Impact of ASC-CON Repeal on Ambulatory Surgical Centers per 100,000 State Population**

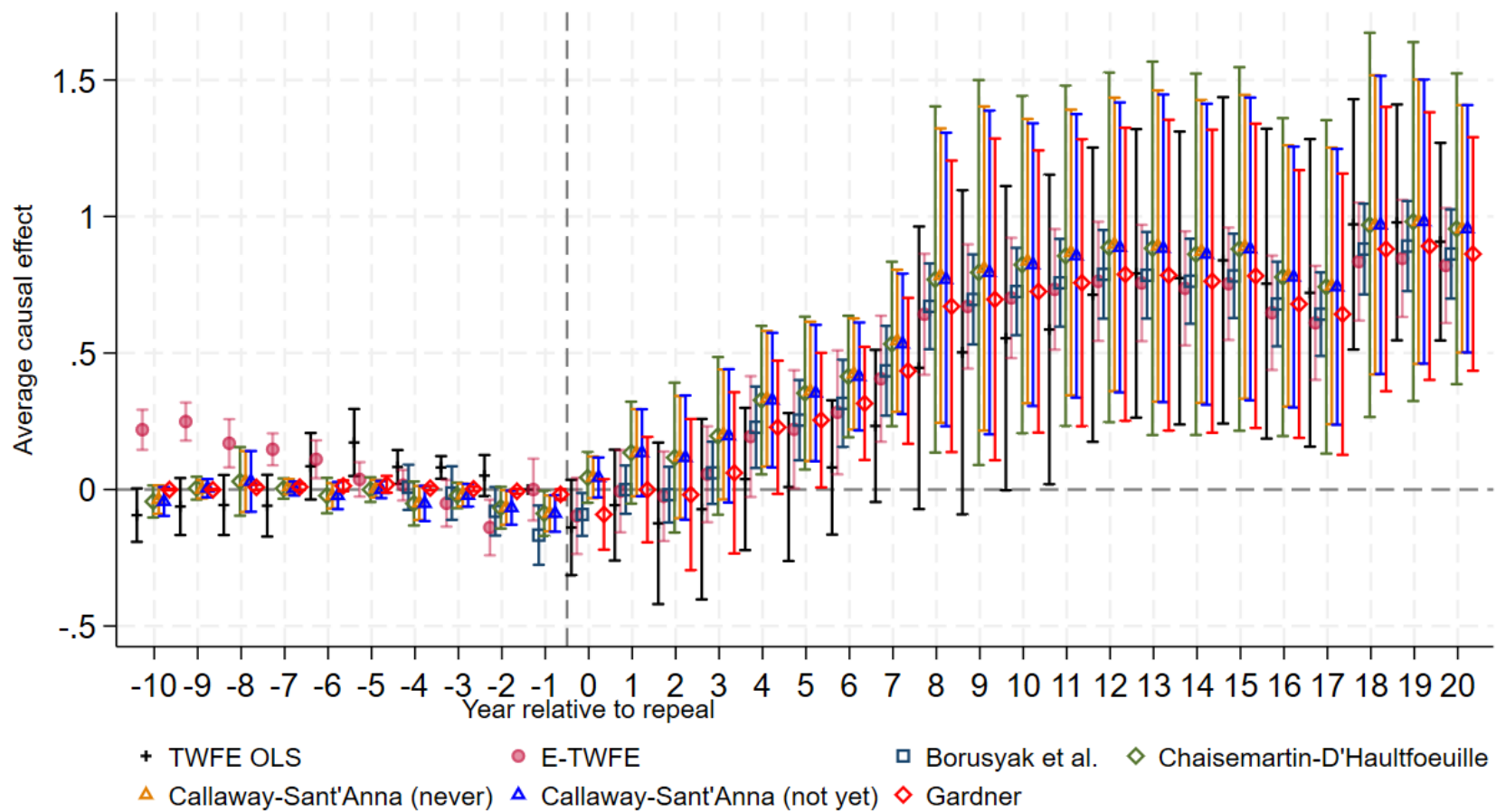

**Table 2. Statewide Effect of ASC-CON Repeal on Ambulatory Surgical Centers per 100,000 State Population**

|  | (1) TWFE | (2) Wooldridge E-TWFE | (3) Borusyak | (4) Callaway Sant'Anna (never) | (5) Callaway Sant'Anna (not yet) | (6) Gardner | (7) Chaisemartin & D'Haultfœuille |
|---|---|---|---|---|---|---|---|
| Estimated Treatment Effect | 0.536** | 0.553*** | 0.553*** | 0.522*** | 0.523*** | 0.553** | 0.523*** |
|  | (0.179) | (0.106) | (0.098) | (0.124) | (0.122) | (0.174) | (0.137) |
| R2 | 0.884 | 0.894 |  |  |  |  |  |
| N | 1,189 | 1,189 | 1,189 | 1,189 | 1,189 | 1,189 | 1,189 |

Note: This table presents the estimated effects of the Certificate of Need repeal on the number of ambulatory surgical centers (ASCs) per 100,000 state population. Column (1) is estimated with Ordinary Least Squares and state and year fixed effects (TWFE). Columns (2)-(7) present aggregated Average Treatment Effects on the Treated (ATT's) from the robust difference-in-difference estimators proposed by Wooldridge (2021), Borusyak et al. (2021), Callaway and Sant'Anna (2021), Gardner (2022), and de Chaisemartin and D'Haultfœuille (2020) using data from 1991-2019 for 40 states. All regressions are estimated with state and year-fixed effects but without other controls. Standard errors clustered at the state level are in parentheses. R-squared is only available for Columns (1) and (2). $^{*} p < 0.05$, $^{**} p < 0.01$, $^{***} p < 0.001$

**Table 3. Rural Effect of ASC-CON Repeal on Ambulatory Surgical Centers per 100,000 State Population**

|  | (1) TWFE | (2) Wooldridge E-TWFE | (3) Borusyak | (4) Callaway Sant'Anna (never) | (5) Callaway Sant'Anna (not yet) | (6) Gardner | (7) Chaisemartin & D'Haultfœuille |
|---|---|---|---|---|---|---|---|
| Estimated Treatment Effect | 0.375 | 0.499*** | 0.499*** | 0.610*** | 0.603*** | 0.499** | 0.603** |
|  | (0.228) | (0.075) | (0.070) | (0.177) | (0.177) | (0.182) | (0.185) |
| R2 | 0.807 | 0.842 | - | - | - | - | - |
| N | 1,073 | 1,073 | 1,073 | 1,073 | 1,073 | 1,073 | 1,073 |

Note: This table presents the estimated effects of the Certificate of Need repeal on the number of rural ambulatory surgical centers (ASCs) per 100,000 rural population. No rural counties exist in Delaware, New Jersey, Rhode Island, and the District of Columbia. Column (1) is estimated with Ordinary Least Squares and state and year fixed effects (TWFE). Columns (2)-(7) present aggregated Estimated Treatment Effects on the Treated (ATT's) from the robust difference-in-difference estimators proposed by Wooldridge (2021), Borusyak et al. (2021), Callaway and Sant'Anna (2021), Gardner (2022), and de Chaisemartin and D'Haultfœuille (2020) using data from 1991-2019 for 36 states. All regressions are estimated with state and year-fixed effects but without other controls. Standard errors clustered at the state level are in parentheses. R-squared is only available for Columns (1) and (2). $^{*} p < 0.05$, $^{**} p < 0.01$, $^{***} p < 0.001$

**Table 4. Effect of ASC-CON Repeal on Rural Hospital Closures per 100,000 State Population**

|  | (1) Closures | (2) Closures | (3) Closures | (4) Closures | (5) Closures |
|---|---|---|---|---|---|
| ASC-CON Repeal | -0.052 | -0.047 | -0.007 | -0.026 | -0.042 |
|  | (0.070) | (0.066) | (0.068) | (0.084) | (0.093) |
| Rural Population, Baseline |  | ✓ | ✓ | ✓ | ✓ |
| Unemployment Rate |  |  | ✓ | ✓ | ✓ |
| Rural Population, % Change |  |  |  | ✓ | ✓ |
| Elderly (65+), % Change |  |  |  | ✓ | ✓ |
| Hispanic, % Change |  |  |  | ✓ | ✓ |
| Black, % Change |  |  |  | ✓ | ✓ |
| Adults Diagnosed Diabetes + Lung Cancer, % Change |  |  |  |  | ✓ |
| R2 | 0.010 | 0.068 | 0.157 | 0.333 | 0.350 |
| N | 37 | 37 | 37 | 37 | 37 |

Note: The dependent variable is the number of Rural Hospital Closures per 100,000 Rural population in the state. No rural counties exist in Delaware, New Jersey, Rhode Island, and the District of Columbia. Column (2) controls for the percentage of the rural state population in 2005 and the average unemployment rate in the state over the 2005-2019 period. Column (3) also controls for the percent change in the rural population between 2005 and 2019. Column (4) also controls for the percent changes in the Elderly (65+), Hispanic, and Black populations in the state over the 2005-2019 period. Column (5) also controls for the percent change in the age-adjusted rate of adults (18+) diagnosed with diabetes and lung cancer from 2005-2016 (these series were discontinued in 2016). Standard errors are clustered by state in parentheses. $^+ p < 0.1$, $^* p < 0.05$, $^{**} p < 0.01$, $^{***} p < 0.001$

**Table 5. Effect of ASC-CON Repeal on Closed Hospital Beds per 100,000 State Population in Rural Areas**

| | (1)<br>Beds | (2)<br>Beds | (3)<br>Beds | (4)<br>Beds | (5)<br>Beds |
|---|---|---|---|---|---|
| ASC-CON Repeal | 0.730 | 0.837 | 2.393 | 1.102 | 0.971 |
| | (3.524) | (3.464) | (3.424) | (3.984) | (4.232) |
| Rural Population, Baseline | | ✓ | ✓ | ✓ | ✓ |
| Unemployment Rate | | | ✓ | ✓ | ✓ |
| Rural Population, % Change | | | | ✓ | ✓ |
| Elderly (65+), % Change | | | | ✓ | ✓ |
| Hispanic, % Change | | | | ✓ | ✓ |
| Black, % Change | | | | ✓ | ✓ |
| Adults Diagnosed Diabetes + Lung Cancer, % Change | | | | | ✓ |
| R2 | 0.001 | 0.022 | 0.125 | 0.336 | 0.336 |
| N | 37 | 37 | 37 | 37 | 37 |

Note: The dependent variable is the number of beds lost to Rural Hospital Closures per 100,000 Rural population in the state. No rural counties exist in Delaware, New Jersey, Rhode Island, and the District of Columbia. Column (2) controls for the percentage of the state population that was rural in 2005 and the average unemployment rate in the state over the 2005-2019 period. Column (3) also controls for the percent change in the rural population between 2005 and 2019. Column (4) also controls for the percent changes in the Elderly (65+), Hispanic, and Black populations in the state over the 2005-2019 period. Column (5) also controls for the percent change in the age-adjusted rate of adults (18+) diagnosed with diabetes and lung cancer from 2005-2016 (these series were discontinued in 2016). Standard errors are clustered by state in parentheses. $^+ p < 0.1$, $^* p < 0.05$, $^{**} p < 0.01$, $^{***} p < 0.001$

Table 6. Effect of ASC-CON Repeal on Hospital Service Reductions per 100,000 State Population in Rural Areas

| | (1) Service Reductions | (2) Service Reductions | (3) Service Reductions | (4) Service Reductions | (5) Service Reductions |
|---|---|---|---|---|---|
| ASC-CON Repeal | -0.091[+] | -0.085[+] | -0.079 | -0.070 | -0.107 |
| | (0.048) | (0.047) | (0.061) | (0.087) | (0.100) |
| Rural Population, Baseline | | ✓ | ✓ | ✓ | ✓ |
| Unemployment Rate | | | ✓ | ✓ | ✓ |
| Rural Population, % Change | | | | ✓ | ✓ |
| Elderly (65+), % Change | | | | ✓ | ✓ |
| Hispanic, % Change | | | | ✓ | ✓ |
| Black, % Change | | | | ✓ | ✓ |
| Adults Diagnosed Diabetes + Lung Cancer, % Change | | | | | ✓ |
| R2 | 0.021 | 0.077 | 0.079 | 0.143 | 0.201 |
| N | 37 | 37 | 37 | 37 | 37 |

Note: The dependent variable is the number of Rural Hospital Service Reductions per 100,000 Rural population in the state. No rural counties exist in Delaware, New Jersey, Rhode Island, and the District of Columbia. Column (2) controls for the percentage of the state population that was rural in 2005 and the average unemployment rate in the state over the 2005-2019 period. Column (3) also controls for the percent change in the rural population between 2005 and 2019. Column (4) also controls for the percent changes in the Elderly (65+), Hispanic, and Black populations in the state over the 2005-2019 period. Column (5) also controls for the percent change in the age-adjusted rate of adults (18+) diagnosed with diabetes and lung cancer from 2005-2016 (these series were discontinued in 2016). Standard errors are clustered by state in parentheses. [+] $p < 0.1$, [*] $p < 0.05$, [**] $p < 0.01$, [***] $p < 0.001$

**Table 7. Effect of ASC-CON Repeal Number of Beds Lost Due to Rural Hospital Service Reductions per 100,000 State Population in Rural Areas**

|  | (1) Beds | (2) Beds | (3) Beds | (4) Beds | (5) Beds |
|---|---|---|---|---|---|
| ASC-CON Repeal | -7.332* | -6.853* | -8.384+ | -10.360 | -12.342+ |
|  | (3.264) | (3.078) | (4.775) | (6.757) | (7.282) |
| Rural Population, Baseline |  | ✓ | ✓ | ✓ | ✓ |
| Unemployment Rate |  |  | ✓ | ✓ | ✓ |
| Rural Population, % Change |  |  |  | ✓ | ✓ |
| Elderly (65+), % Change |  |  |  | ✓ | ✓ |
| Hispanic, % Change |  |  |  | ✓ | ✓ |
| Black, % Change |  |  |  | ✓ | ✓ |
| Adults Diagnosed Diabetes + Lung Cancer, % Change |  |  |  |  | ✓ |
| R2 | 0.022 | 0.085 | 0.100 | 0.183 | 0.210 |
| N | 37 | 37 | 37 | 37 | 37 |

Note: The dependent variable is the number of beds lost due to Rural Hospital Service Reductions per 100,000 Rural population in the state. No rural counties exist in Delaware, New Jersey, Rhode Island, and the District of Columbia. Column (2) controls for the percentage of the state population that was rural in 2005 and the average unemployment rate in the state over the 2005-2019 period. Column (3) also controls for the percent change in the rural population between 2005 and 2019. Column (4) also controls for the percent changes in the Elderly (65+), Hispanic, and Black populations in the state over the 2005-2019 period. Column (5) also controls for the percent change in the age-adjusted rate of adults (18+) diagnosed with diabetes and lung cancer from 2005-2016 (these series were discontinued in 2016). Standard errors are clustered by state in parentheses. $^+ p < 0.1$, $^* p < 0.05$, $^{**} p < 0.01$, $^{***} p < 0.001$

**Appendix**

All Figures and Tables in the Appendix define the control group as including states that never had a CON during the sample period and states that always had a CON during the sample period.

**Figure A1. Statewide Event-Study Results: Impact of ASC-CON Repeal on Ambulatory Surgical Centers per 100,000 State Population**

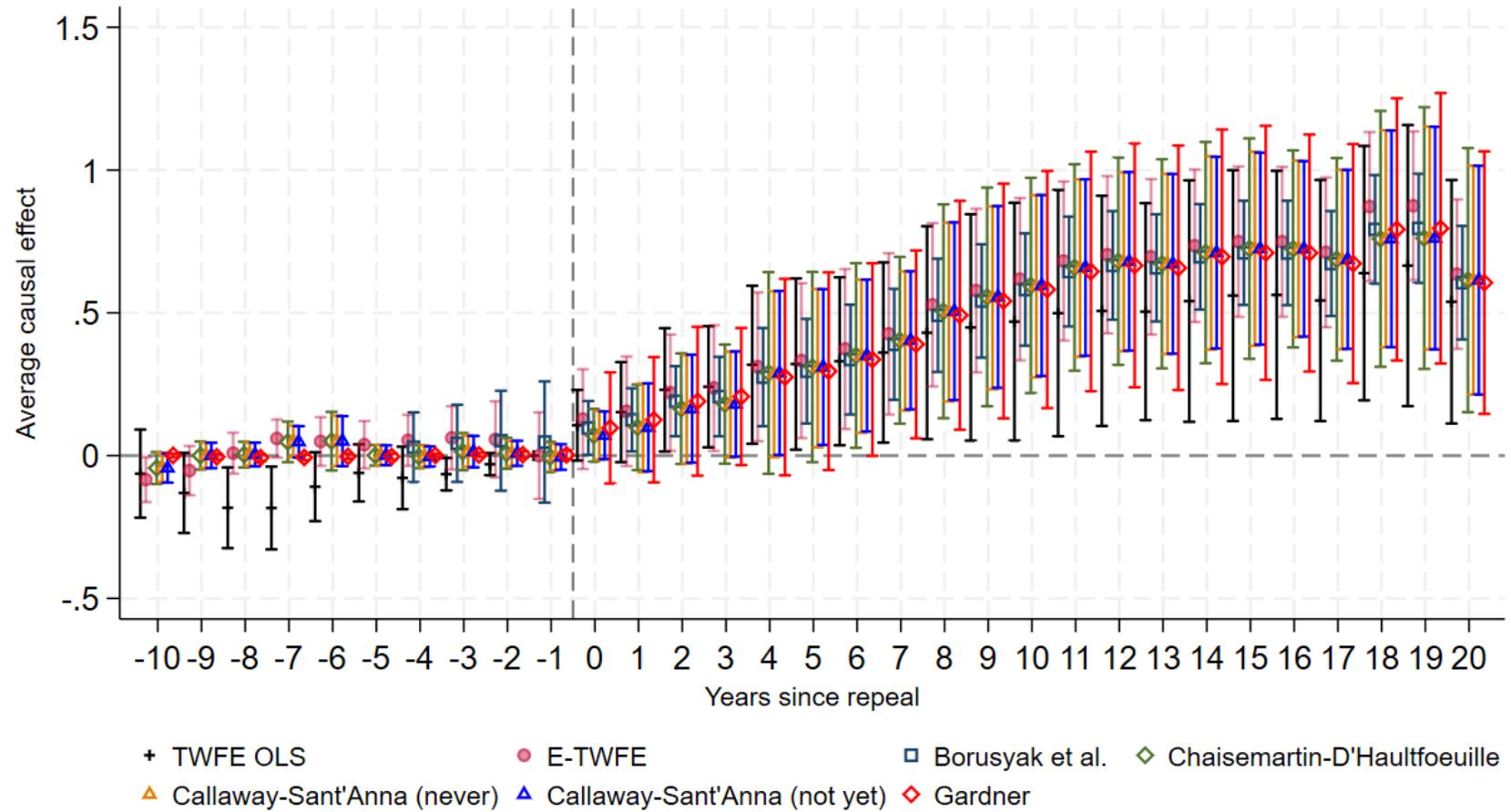

**Figure A2. Rural Event-Study Results: Impact of ASC-CON Repeal on Ambulatory Surgical Centers per 100,000 State Population**

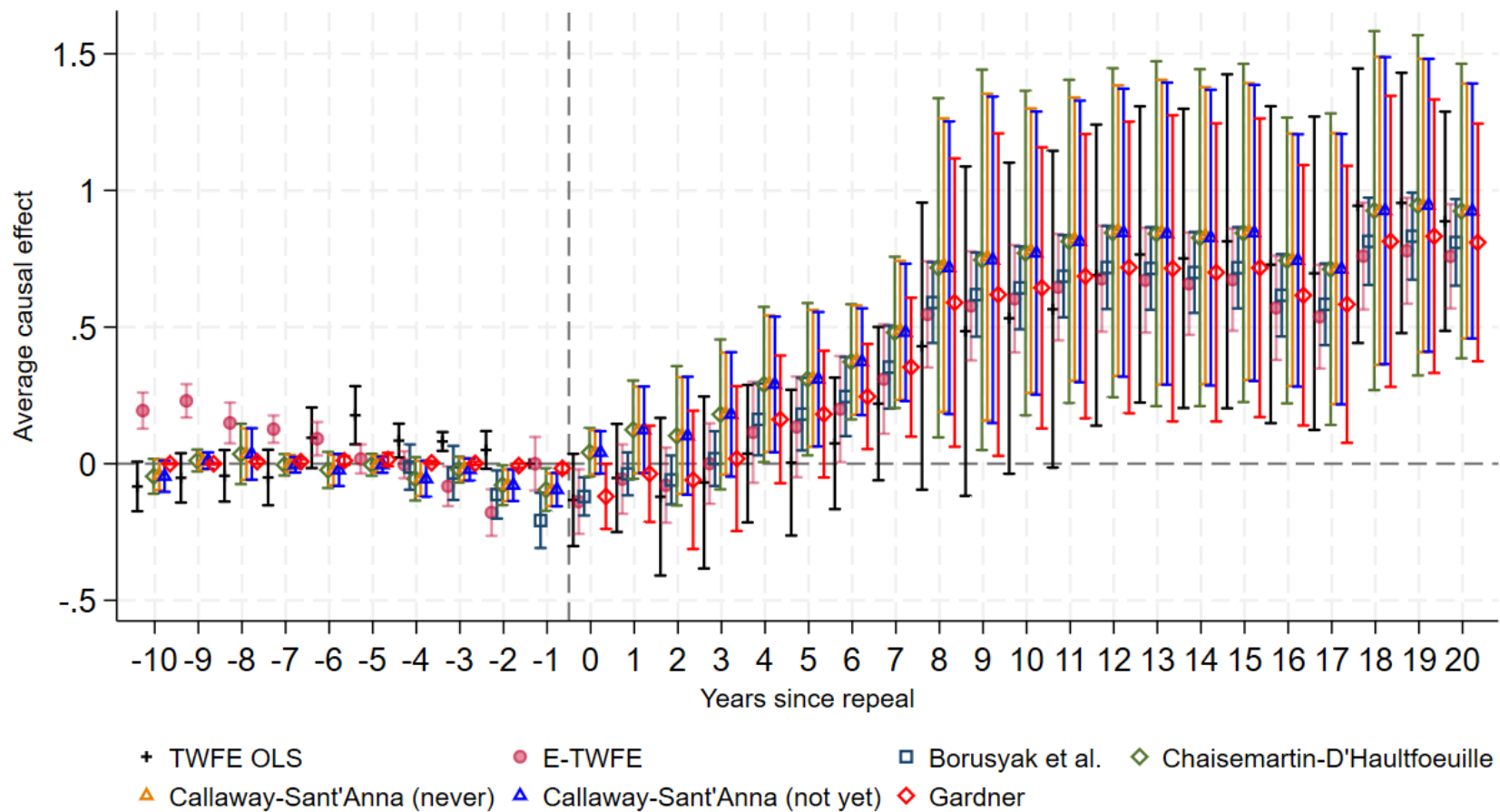

**Table A1. Statewide Effect of ASC-CON Repeal on Ambulatory Surgical Centers per 100,000 State Population**

|  | (1) TWFE | (2) Wooldridge E-TWFE | (3) Borusyak et al. | (4) Callaway Sant'Anna (never) | (5) Callaway Sant'Anna (not yet) | (6) Gardner | (7) Chaisemartin & D'Haultfœuille |
|---|---|---|---|---|---|---|---|
| Estimated Treatment Effect | 0.471** (0.170) | 0.476*** (0.092) | 0.476*** (0.086) | 0.478*** (0.121) | 0.480*** (0.120) | 0.476** (0.167) | 0.480*** (0.126) |
| R2 | 0.877 | 0.885 | - | - | - | - | - |
| N | 1,479 | 1,479 | 1,479 | 1,479 | 1,479 | 1,479 | 1,479 |

Note: This table presents the estimated effects of the Certificate of Need repeal on the number of ambulatory surgical centers (ASCs) per 100,000 state population. Column (1) is estimated with Ordinary Least Squares and state and year fixed effects (TWFE). Columns 2-7 present aggregated Average Treatment Effects on the Treated (ATT's) from the robust difference-in-difference estimators proposed by Wooldridge (2021), Borusyak et al. (2021), Callaway and Sant'Anna (2021), Gardner (2022), and de Chaisemartin and D'Haultfœuille (2020) using all available data from 1991-2019 for 51 states. All regressions are estimated with state and year-fixed effects but without other controls. R-squared is only available for Columns (1) and (2). Standard errors clustered at the state level in parentheses. R-squared is only available for Columns (1) and (2). $^+ p < 0.1$, $^* p < 0.05$, $^{**} p < 0.01$, $^{***} p < 0.001$

**Table A2. Rural Effect of ASC-CON Repeal on Ambulatory Surgical Centers per 100,000 State Population**

|  | (1) TWFE | (2) Wooldridge E-TWFE | (3) Borusyak et al. | (4) Callaway Sant'Anna (never) | (5) Callaway Sant'Anna (not yet) | (6) Gardner | (7) Chaisemartin & D'Haultfœuille |
|---|---|---|---|---|---|---|---|
| Estimated Treatment Effect | 0.327 (0.220) | 0.438*** (0.069) | 0.438*** (0.066) | 0.573** (0.176) | 0.569** (0.176) | 0.438* (0.178) | 0.569** (0.190) |
| R2 | 0.793 | 0.819 | - | - | - | - | - |
| N | 1,363 | 1,363 | 1,363 | 1,363 | 1,363 | 1,363 | 1,363 |

Note: This table presents the estimated effects of the Certificate of Need repeal on the number of rural ambulatory surgical centers (ASCs) per 100,000 rural population. Column (1) is estimated with Ordinary Least Squares and state and year fixed effects (TWFE). Columns 2-7 present aggregated Average Treatment Effects on the Treated (ATT's) from the robust difference-in-difference estimators proposed by Wooldridge (2021), Borusyak et al. (2021), Callaway and Sant'Anna (2021), Gardner (2022), and de Chaisemartin and D'Haultfœuille (2020) using all available data from 1991-2019 for 47 states. No rural counties exist in Delaware, New Jersey, Rhode Island, and the District of Columbia. All regressions are estimated with state and year-fixed effects but without other controls. Standard errors clustered at the state level are in parentheses. R-squared is only available for Columns (1) and (2). $^+ p < 0.1$, $^* p < 0.05$, $^{**} p < 0.01$, $^{***} p < 0.001$

**Table A4. Effect of ASC-CON Repeal on Rural Hospital Closures per 100,000 State Population**

|  | (1) Hospital Closures | (2) Hospital Closures | (3) Hospital Closures | (4) Hospital Closures | (5) Hospital Closures |
|---|---|---|---|---|---|
| ASC-CON Repeal | -0.051 (0.069) | -0.049 (0.063) | -0.021 (0.064) | -0.028 (0.075) | -0.013 (0.074) |
| Rural Population, Baseline |  | ✓ | ✓ | ✓ | ✓ |
| Unemployment Rate |  |  | ✓ | ✓ | ✓ |
| Rural Population, % Change |  |  |  | ✓ | ✓ |
| Elderly (65+), % Change |  |  |  | ✓ | ✓ |
| Hispanic, % Change |  |  |  | ✓ | ✓ |
| Black, % Change |  |  |  | ✓ | ✓ |
| Adults Diagnosed Diabetes + Lung Cancer, % Change |  |  |  |  | ✓ |
| R2 | 0.007 | 0.118 | 0.210 | 0.244 | 0.257 |
| N | 47 | 47 | 47 | 47 | 47 |

Note: The dependent variable is the number of Rural Hospital Closures per 100,000 Rural population in the state. No rural counties exist in Delaware, New Jersey, Rhode Island, and the District of Columbia. Column (2) controls for the percentage of the rural state population in 2005 and the average unemployment rate in the state over the 2005-2019 period. Column (3) also controls for the percent change in the rural population between 2005 and 2019. Column (4) also controls for the percent changes in the Elderly (65+), Hispanic, and Black populations in the state over the 2005-2019 period. Column (5) also controls for the percent change in the age-adjusted rate of adults (18+) diagnosed with diabetes and lung cancer from 2005-2016 (these series were discontinued in 2016). Standard errors are clustered by state in parentheses. $^+ p < 0.1$, $^* p < 0.05$, $^{**} p < 0.01$, $^{***} p < 0.001$

**Table A5. Effect of ASC-CON Repeal on Closed Hospital Beds per 100,000 State Population in Rural Areas**

| | (1) Beds | (2) Beds | (3) Beds | (4) Beds | (5) Beds |
|---|---|---|---|---|---|
| ASC-CON Repeal | 0.432 | 0.489 | 1.650 | 0.862 | 1.784 |
| | (3.511) | (3.359) | (3.266) | (3.678) | (3.643) |
| Rural Population, Baseline | | ✓ | ✓ | ✓ | ✓ |
| Unemployment Rate | | | ✓ | ✓ | ✓ |
| Rural Population, % Change | | | | ✓ | ✓ |
| Elderly (65+), % Change | | | | ✓ | ✓ |
| Hispanic, % Change | | | | ✓ | ✓ |
| Black, % Change | | | | ✓ | ✓ |
| Adults Diagnosed Diabetes + Lung Cancer, % Change | | | | | ✓ |
| R2 | 0.000 | 0.080 | 0.181 | 0.227 | 0.259 |
| N | 47 | 47 | 47 | 47 | 47 |

Note: The dependent variable is the number of beds lost to Rural Hospital Closures per 100,000 Rural population in the state. No rural counties exist in Delaware, New Jersey, Rhode Island, and the District of Columbia. Column (2) controls for the percentage of the rural state population in 2005 and the average unemployment rate in the state over the 2005-2019 period. Column (3) also controls for the percent change in the rural population between 2005 and 2019. Column (4) also controls for the percent changes in the Elderly (65+), Hispanic, and Black populations in the state over the 2005-2019 period. Column (5) also controls for the percent change in the age-adjusted rate of adults (18+) diagnosed with diabetes and lung cancer from 2005-2016 (these series were discontinued in 2016). Standard errors are clustered by state in parentheses. $^+ p < 0.1$, $^* p < 0.05$, $^{**} p < 0.01$, $^{***} p < 0.001$

**Table A6. Effect of ASC-CON Repeal on Hospital Service Reductions per 100,000 State Population in Rural Areas**

|  | (1) Service Reductions | (2) Service Reductions | (3) Service Reductions | (4) Service Reductions | (5) Service Reductions |
|---|---|---|---|---|---|
| ASC-CON Repeal | -0.097* | -0.096* | -0.098+ | -0.083 | -0.095 |
|  | (0.043) | (0.044) | (0.052) | (0.070) | (0.073) |
| Rural Population, Baseline |  | ✓ | ✓ | ✓ | ✓ |
| Unemployment Rate |  |  | ✓ | ✓ | ✓ |
| Rural Population, % Change |  |  |  | ✓ | ✓ |
| Elderly (65+), % Change |  |  |  | ✓ | ✓ |
| Hispanic, % Change |  |  |  | ✓ | ✓ |
| Black, % Change |  |  |  | ✓ | ✓ |
| Adults Diagnosed Diabetes + Lung Cancer, % Change |  |  |  |  | ✓ |
| R2 | 0.019 | 0.037 | 0.037 | 0.165 | 0.172 |
| N | 47 | 47 | 47 | 47 | 47 |

Note: The dependent variable is the number of Rural Hospital Service Reductions per 100,000 Rural population. No rural counties exist in Delaware, New Jersey, Rhode Island, and the District of Columbia. Column (2) controls for the percentage of the rural state population in 2005 and the average unemployment rate in the state over the 2005-2019 period. Column (3) also controls for the percent change in the rural population between 2005 and 2019. Column (4) also controls for the percent changes in the Elderly (65+), Hispanic, and Black populations in the state over the 2005-2019 period. Column (5) also controls for the percent change in the age-adjusted rate of adults (18+) diagnosed with diabetes and lung cancer from 2005-2016 (these series were discontinued in 2016). Standard errors are clustered by state in parentheses. + $p < 0.1$, * $p < 0.05$, ** $p < 0.01$, *** $p < 0.001$

**Table A7. Effect of ASC-CON Repeal on Facility Size in Service Reductions per 100,000 State Population in Rural Areas**

| | (1) Beds | (2) Beds | (3) Beds | (4) Beds | (5) Beds |
|---|---|---|---|---|---|
| ASC-CON Repeal | -6.407* | -6.318* | -6.994* | -8.479+ | -9.281+ |
| | (2.530) | (2.572) | (3.388) | (4.849) | (4.830) |
| Rural Population, Baseline | | ✓ | ✓ | ✓ | ✓ |
| Unemployment Rate | | | ✓ | ✓ | ✓ |
| Rural Population, % Change | | | | ✓ | ✓ |
| Elderly (65+), % Change | | | | ✓ | ✓ |
| Hispanic, % Change | | | | ✓ | ✓ |
| Black, % Change | | | | ✓ | ✓ |
| Adults Diagnosed Diabetes + Lung Cancer, % Change | | | | | ✓ |
| R2 | 0.017 | 0.065 | 0.073 | 0.133 | 0.139 |
| N | 47 | 47 | 47 | 47 | 47 |

Note: The dependent variable is the number of beds in Rural Hospital Service Reductions per 100,000 rural population in the state. No rural counties exist in Delaware, New Jersey, Rhode Island, and the District of Columbia. Column (2) controls for the percentage of the state population that was rural in 2005 and the average unemployment rate in the state over the 2005-2019 period. Column (3) also controls for the percent change in the rural population between 2005 and 2019. Column (4) also controls for the percent changes in the Elderly (65+), Hispanic, and Black populations in the state over the 2005-2019 period. Column (5) also controls for the percent change in the age-adjusted rate of adults (18+) diagnosed with diabetes and lung cancer from 2005-2016 (these series were discontinued in 2016). Standard errors are clustered by state in parentheses. $^+ p < 0.1$, $^* p < 0.05$, $^{**} p < 0.01$, $^{***} p < 0.001$